\documentclass[sigconf]{acmart}

\AtBeginDocument{%
  \providecommand\BibTeX{{%
    \normalfont B\kern-0.5em{\scshape i\kern-0.25em b}\kern-0.8em\TeX}}}

\setcopyright{acmcopyright}
\copyrightyear{2018}
\acmYear{2018}
\acmDOI{XXXXXXX.XXXXXXX}

\acmConference[Conference acronym 'XX]{Make sure to enter the correct
  conference title from your rights confirmation emai}{June 03--05,
  2018}{Woodstock, NY}
\acmPrice{15.00}
\acmISBN{978-1-4503-XXXX-X/18/06}

\usepackage{color}
\usepackage{listings}
\usepackage[ruled,linesnumbered]{algorithm2e}

\usepackage{mathtools}
\usepackage{array}
\usepackage{stmaryrd}
\usepackage{amsmath}
\usepackage{amsfonts}
\usepackage{tikz}
\usepackage{pgfplots}
\usepackage{enumitem}
\usepackage[font=singlespacing]{caption}

\newtheorem{problem}{Problem}[section]

\usepackage{listings}
\usepackage{xcolor}
\usepackage{proof}

\definecolor{codegreen}{rgb}{0,0.6,0}
\definecolor{codegray}{rgb}{0.5,0.5,0.5}
\definecolor{codepurple}{rgb}{0.58,0,0.82}
\definecolor{backcolour}{rgb}{0.95,0.95,0.92}

\lstdefinestyle{mystyle}{
    commentstyle=\color{codegreen},
    keywordstyle=\color{magenta},
    numberstyle=\tiny\color{codegray},
    stringstyle=\color{codepurple},
    basicstyle=\ttfamily\footnotesize,
    breakatwhitespace=false,         
    breaklines=true,                 
    captionpos=b,                    
    keepspaces=true,                 
    numbers=left,                    
    numbersep=5pt,                  
    showspaces=false,                
    showstringspaces=false,
    showtabs=false,                  
    tabsize=2
}

\begin{document}

\title{A Data Source Dependency Analysis Framework for Large Scale Data Science Projects} 


\author{Laurent Bou\'e}
\email{laurent.boue@microsoft.com}
\affiliation{%
  \institution{Microsoft}
  \country{Israel}
}
\author{Pratap Kunireddy}
\email{pkunireddy@microsoft.com}
\affiliation{%
  \institution{Microsoft}
  \country{India}
}
\author{Pavle Suboti\'{c}}
\email{pavlesubotic@microsoft.com}
\affiliation{%
  \institution{Microsoft}
  \country{Serbia}
}

\renewcommand{\shortauthors}{et al.}

\begin{abstract}
Dependency hell is a well-known pain point in the development of large software projects and machine learning (ML) code bases are not 
immune from it.  In fact, ML applications suffer from an additional form, namely,  \emph{data source dependency hell}.  This term refers to the central 
role played by data and its unique quirks that often lead to unexpected failures of ML models which cannot be explained by code changes.  In this paper, we 
present an automated dependency mapping framework that allows MLOps engineers to 
monitor the whole dependency map of their models in a fast paced engineering environment and thus mitigate ahead of time the consequences of any data source changes (e.g., re-train model, ignore data, set default data etc.).  Our system is based on a unified and 
generic approach, employing techniques from static analysis, from which data sources can 
be identified reliably for any type of dependency on a wide range of source languages and artifacts. The dependency mapping framework is exposed as a 
REST web API where the only input is the path to the Git 
repository hosting the code base.  Currently used by MLOps engineers at Microsoft, we 
expect such dependency map APIs to be adopted more widely by MLOps engineers in the future.
\end{abstract}




\maketitle

\section{Introduction}

With the widespread adoption of machine learning (ML) models, it is common  for large data science organizations to 
run hundreds of mission-critical workloads daily.  Any failures or delays negatively affect stakeholders who do not receive the 
information they depend on to generate business value and deliver results to their customers.  In this context, a new discipline that aims to streamline and standardize all the engineering pieces necessary to land production-level quality ML models at scale
has emerged under the name of MLOps.  Although 
similar in spirit and execution to DevOps, MLOps engineers must also deal with a new set of challenges associated with 
data-centric (instead of code-centric) projects~\cite{9720902,SEforML}. 
MLOps includes incident management, data quality processes (model / data drift checks), data contracts, operationalization / support of models and other responsibilities to automate the lifecycle and continuous delivery of high-performing models in production.
All this must be accounted for, while ensuring scale, latency, speed and optimal costs. 

In particular, the topic of \emph{data source dependencies} is central to MLOps.  As models evolve and grow in scope, the overall MLOps 
architecture rapidly becomes overwhelmed by a new form of \emph{data source dependency hell} where, instead of software packages’ versions, 
MLOps engineers now must try to resolve models’ data source dependencies.  Generally, this task is even more laborious because 
data dependencies tend to span across organizational boundaries and originate from a diversity of data stores. It is essential to note that while code issues surface occasionally, data issues arise frequently and rather unpredictably.   Unfortunately, important data source announcements (such as delays or issues of any kind) which are often communicated via distribution list 
emails commonly go unnoticed by model owners thereby resulting in model failures and ultimately negative customer impact.  Even when data 
teams assign a specific role to manually track all such announcements, the task of deciphering which models are impacted remains daunting due 
to the inherently iterative nature of data science models and to their complex dependency graphs 
(with transitive and model-model structures).


In this paper, we present an automated dependency mapping
framework to extract data source dependencies from ML models. It allows subscribed systems to automatically notify a model owner if, for example, 
a dependent table schema has changed, contains corrupted data or any other type of modification that is likely 
to affect the ML model.  Upon receipt of the notification, the model owner can quickly take the necessary mitigating actions (e.g., re-train model, ignore data, change data source etc.) so that model consumers are not affected.  To adequately perform automated data source mapping, our technique employs static analysis on a set of connected code artifacts referred to as \emph{activity graphs}. The advantage of our approach is that we can compute data source dependencies quickly and with ease thus avoiding run-time complications (e.g., security configurations, storage costs etc.). Moreover, our framework is extensible in that it supports a wide range of activity artifacts including queries, scripts and notebooks without requiring modifications. We have implemented our automated dependency mapping framework and exposed it as a web API service.  Our implementation is deployed within Microsoft Cloud Data Sciences (MCDS) where it is used to mitigate and prevent data source related incidents ahead of time. 

We summarize our contributions as follows:

\begin{itemize}
    \item We present a novel dependency mapping framework that extracts data sources from activity graphs using static analysis.
    \item We present an implementation of our technique and deployment in real-world industrial use cases. 
    \item We evaluate our framework and show its utility on real-world benchmarks.
\end{itemize}

\section{Problem Definition}
\label{sec:prob}

\begin{figure}
\centering
\includegraphics[width=0.5\textwidth]{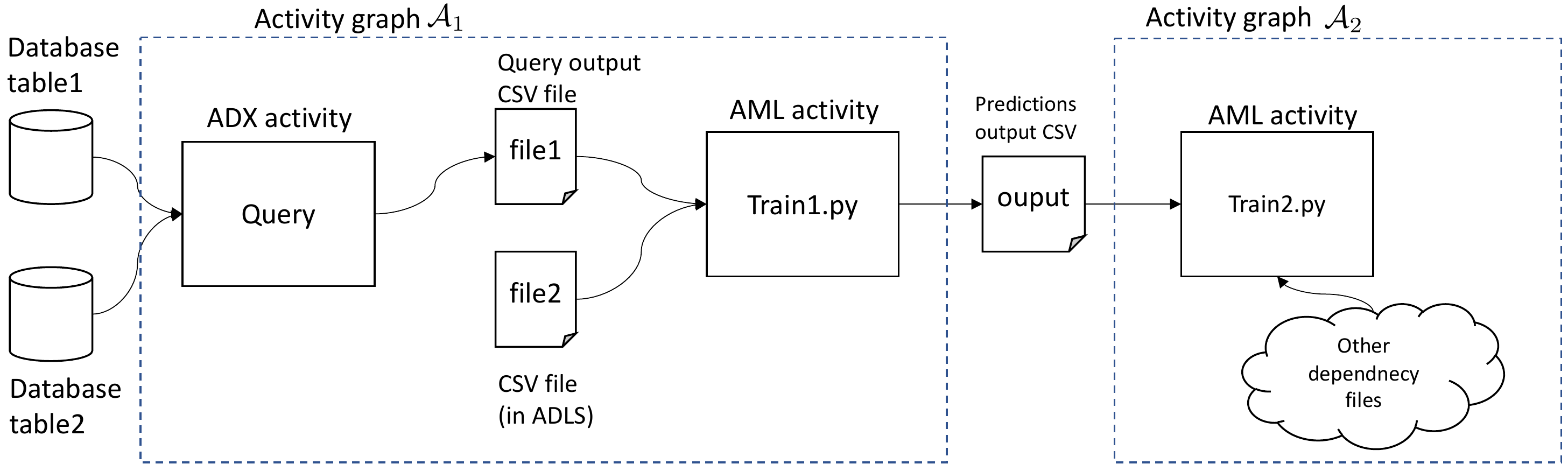}
\caption{Dependency Mapping Example\label{fig:motivation}}
\end{figure}

The need to automate dependency mappings comes from the typical volumes as well as time scales associated with data sources.  For instance at the time of writing, a modestly-sized data science organization deployed on our framework needs to manage approx. 300 activities, 25 databases, 250 tables and 2,500 columns.  Additionally, models are continuously updated and modified by model owners so that data sources routinely appear or disappear on an hourly basis.  On top of this, models and data sources often display complex transitive dependencies with one another.  In this context, it is infeasible to manually track data source dependencies.  Automatic monitoring also allows data science organizations to reduce their rates of missed Service Level Agreements (SLAs) with their stakeholders by providing a faster path to identify root cause reasons for model failures and resolve incidents before they are even noticed by downstream stakeholders.

A typical ML workload is constructed as one or several \emph{activity graphs}. An activity graph is a set of Azure Data Factory (ADF) \emph{activities} such as Azure Data Explorer (ADX) or Azure Machine Learning (AML). Activities digest data and produce output data. Hence, activities are connected via intermediate data. We classify data into two categories, namely, initial data sources (i.e., raw data from a database or file etc.) and derived data source (i.e., output from another activity).

Activities perform a well-defined unit of work such as control flow operations, data mapping operations and many others using programming logic i.e., scripts, database queries etc.  Normally, ADX activities are responsible for data querying and AML activities for training~/ ~inference of ML models. 

We define a data source as a tuple $\langle s, \bar{c} \rangle$ where $s$ is a source symbol i.e., representing a database table name or filename, $\bar{c}$ is the set of columns. We use the notation 
$s^{\bar{c}}$. Typically, an activity graph is defined \emph{per} model and its output is the output of a model e.g., inference. Output from a model i.e., an activity graph, can be an input into another activity graph.

\begin{problem}[Model-Input Dependency Mapping]
\label{prob:prob1}
Given a set of initial data sources and a set of models in activity graphs, determine which initial data sources are used by which models.
\end{problem}

\begin{example}[Motivating Example]
In Figure~\ref{fig:motivation} we describe a small
configuration with two activity graphs $\mathcal{A}_1$ and $\mathcal{A}_2$ (enclosed by the dotted lines).
In the first 
graph, 
we have two activities, namely, an~ADX database query and an~AML Python script. The database query is performed on two database tables and on some set of columns (which we don't define for readability). 
Thus \emph{table1} and \emph{table2} are initial data sources whereas \emph{file1} is a data source derived from the~ADX activity.
Both \emph{file1} (derived data source) along with another \emph{file2} (initial data source) serve as inputs to the~AML activity consisting of a machine learning script \emph{Train1.py} which stores its predictions in an \emph{output} file.
To slightly add to the complexity, the derived data source \emph{output} which is produced by~$\mathcal{A}_1$ is used as input to train another model \emph{Train2.py} in another AML activity which belongs to the second activity graph~$\mathcal{A}_2$.
\end{example}
 
In the next sections we describe how we construct an initial data source map for each model to  
solve Problem~\ref{prob:prob1}.

\section{Dependency Mapping Framework}
In this section we describe our 
dependency mapping algorithm. Our algorithm 
operates on three 
levels: (1) set of connected activity graphs 
(2) a single activity graph
(3) a single activity. 

\begin{algorithm}
\caption{Activity Graph Analysis}\label{alg:agraph}
\SetKwInOut{Input}{Input}
\Input{Activity graph $\mathcal{A}$}
\KwResult{Set of data sources $\zeta$}
$\zeta := \emptyset$\; \label{alg1:init}
$nodes$ := FIFO()\;
$nodes$.push($\mathcal{A}$.start)\;\label{alg1:start}
\While{$nodes$ is not empty}
{
a := $nodes$.pop()\;
I := \textbf{Analyze}(a)\; \label{alg1:analyze}

\If{$I$ reached fixpoint} {
 \Return $\zeta$\;\label{alg1:ret1}
}
\For{$i \in I \wedge \mathcal{A}.derived(i)$}
{\label{alg1:derived}
nodes.push($\mathcal{A}.deps(i)$)\; \label{alg1:prop}
}
$\zeta := \zeta \cup \{i \in I \ | \ source(i) \}$\; \label{alg1:source}
}
\Return $\zeta$\;\label{alg1:ret2}
\end{algorithm}

Algorithm~\ref{alg:agraph} describes our approach for a single activity graph~$\mathcal{A}$. For each $\mathcal{A}$ we assume a start activity $start$, a function $deps$ that given a data source provides a set of activities in $\mathcal{A}$ that produced it as output and a predicate $derived$ which asserts if a data source is derived. In line~\ref{alg1:init}, we initialize the data source set~$\zeta$ to the empty set. We then create a FIFO ($nodes$) and add the starting activity to it. The general idea is that we propagate backwards (line~\ref{alg1:prop}) to other connected activities until we encounter only initial data sources or stop gaining information (reach a fixpoint). Each time we encounter an activity, we analyze the activity statically using the \textbf{Analyze} function (line~\ref{alg1:analyze}). We then proceed to following dependencies (via $deps$) from derived data (line~\ref{alg1:derived}) and update $\zeta$ when initial data sources (line~\ref{alg1:source}) are found. The algorithm terminates when either no more nodes are in the FIFO (line~\ref{alg1:ret2}) or in the case of a cycle, we detect a fixpoint (due to the assumed monotonicity of \textbf{Analyze}) in line~\ref{alg1:ret1}.

\begin{example}[Motivating Example Cont.]
 Consider our motivating example in Figure~\ref{fig:motivation}. Suppose we analyze the first activity graph. We first process \emph{Train1.py} and 
 have $\zeta = \{file2^{age}\}$ and mark $file1$ as a derived 
 data source. We then add the query activity to our $nodes$ and find as 
 a result that we have $\zeta = \{file2^{age}, table1^{loc}, table2^{name}\}$ assuming the query selects columns $loc$ and $name$.
\end{example}

For the case of several activity graphs we 
 apply Algorithm~\ref{alg:agraph} to individual 
activity graphs and repeatedly make the following inference: \infer{\zeta_B = (\zeta_B - o) \cup \zeta_A }{%
    & o \in O_A
&  o \in \zeta_B
} until a fixpoint is reached. Here, $O_A$ is the set of all outputs of the a model in an activity graph $A$, and $\zeta_B$ is a mapping in an activity graph $B$. This rule states: \emph{if there is a common element between an output of activity graph A and a data source mapping of activity graph B then there is a transitive mapping to activity graph $B$ for all data sources that are mapped to activity graph $A$, excluding the common element}. 

\begin{example}[Motivating Example Cont.]
Consider our motivating example in Figure~\ref{fig:motivation}. We first build the graph for the model in \emph{Train2.py}. Here $\zeta_{\mathcal{A}_2} = \{output^{\dots}, file2^{name}, \dots\}$. Since $O_{\mathcal{A}_1} = \{output^{\dots}\}$ and $\zeta_{\mathcal{A}_1} = \{file2^{age}, table1^{loc}, table2^{name}\}$ we can then conclude that $\zeta_{\mathcal{A}_2} = \{file2^{age}, table1^{loc}, table2^{name}, \dots\}$.
\end{example}

\section{Dependencies From Source Code}
\label{sec:staticanalysis}
In this section we describe our static analysis 
techniques to implement the \textbf{Analyze} function 
in line~\ref{alg1:analyze} of Algorithm~\ref{alg:agraph} that determines model-data dependencies from 
source code in an activity. 

\subsection{Source Mapping Analysis for Scripts}
\label{sec:scripts}

\begin{figure}
\centering
\begin{lstlisting}[language=Python, xleftmargin=.55in]
...
...
data1 = pd.read_csv("file1.csv")
data2 = pd.read_csv("file2.csv")
X = data1[["loc", "age"]]
y = data2[["target"]]
X_train, X_test, y_train, y_test =
    train_test_split(X, y, ...)
lr = LogisticRegression()
a = lr.fit(X_train, y_train)
y_pred = lr.predict(X_test)
y_pred.to_csv("output.csv")
\end{lstlisting}
\caption{Train1.py Script\label{code:train1}}
\end{figure}
\paragraph{Static Analysis} To statically analyze scripts we perform an abstract interpretation\footnote{Can also be seen as a data flow analysis}. The framework of abstract interpretation~\cite{CC77,Kildall73}, computes an over-approximation $\sigma^\sharp$ of a state $\sigma$ by 
iteratively solving the fixpoint equation $\sigma^\sharp = \sigma^{\sharp}_0 \sqcup \llbracket p \rrbracket^\sharp(\sigma^\sharp)$ using a monotonic interpretation of the abstract semantics
$\llbracket p \rrbracket^\sharp$ for a program $p$, that produces an updated abstract state that is joined ($\sqcup$) with the initial abstract state $\sigma^\sharp_0$. We refer the reader to~\cite{cousot2021principles,khedker2017data} for an in-depth explanation of abstract interpretation and data flow analysis. 

Practically, the above is achieved by converting source code into an
abstract syntax tree (AST) representation and subsequently into a control-flow graph (CFG), which is then analyzed. Given a sequence of statements, the CFG is a directed graph that encodes the 
control flow of the nodes that represent straight-line statements (no branching etc.). We define a CFG as $\langle L, E \rangle$ where 
an edge $(l, st, l') \in E$ reflects the semantics of statement $st$ associated with the CFG edge from 
locations $l$ to  $l'$. The set of variables in all the statements is denoted by $V$ and the set of symbols by $S$. The analysis proceeds to compute a (least) fixpoint solution by processing each statement starting with the entry statement and following the CFG control-flow using a worklist algorithm~\cite{khedker2017data}. 

To instantiate an abstract interpretation for our dependency mapping problem, we  need to define two elements (1) how we represent that abstract computational state and (2) how to define the abstract semantics i.e., rules for how we process each type of statement.

\paragraph{Abstract state} We define an abstract state as a mapping between variables and set of input data sources e.g., $\sigma^\sharp =  v \mapsto \{\dots, s^{\bar{c}}, \dots\} $ where $v \in V$, $s$ is a data source symbol with selected set of columns $\bar{c}$. We define a $\sqcup$ operator on states as a piece wise set union. We denote substitution of a variable $x$ with value $d$ in an abstract state $\sigma^\sharp$ as  $\sigma^\sharp[x \mapsto d]$. Apart from the abstract state, we also introduce a \emph{mapping set} $I$ which contains the data sources that reach a model function. 

\paragraph{Abstract semantics} To define our abstract semantics we introduce an abstract transformer, namely, 
a function $\llbracket st \rrbracket$ parameterized by a 
statement type $st$ that takes as arguments (denoted as $\lambda$ arguments) an abstract state $\sigma^\sharp$  and mapping set $I$. The abstract transformer specifies how we analyze each type of statement  and returns an updated abstract state and mapping set. Moreover, our abstract semantics requires us to detect input and model operation statements.  We thus mark 
the set of input statements (e.g., read\_csv) in a set \emph{Source} and 
the set of model training statements (e.g., fit) in a set \emph{Sink}.  The result of an analysis is a mapping set $I$ which is a set of data sources. Note, by the definition of our problem, we assume only one model per activity graph. 

Below, we outline the key set of rules that govern how input sources are propagated for a given statement.

\begin{enumerate}[leftmargin=*]
\item \textbf{input:}
\begin{align*}
  &\lambda \sigma^\sharp.I.\llbracket y = \text{read}(input) \rrbracket  =
    \begin{array}{l}
     \langle \sigma^\sharp \cup  y \mapsto \{input^{C}\}\rangle, I
    \end{array}\\
  & \text{where read } \in Source,
    C \text{ set of all columns of input}
\end{align*}

\item \textbf{project:}
\begin{align*}
     &\lambda \sigma^\sharp.I.\llbracket y = x.sel[\bar{c}] \rrbracket  =  \sigma^\sharp[y \mapsto \sigma^\sharp(x)^{\bar{c}}], I \\
     & \text{where } \sigma^\sharp(x)^{\bar{c}} \text{ constrains columns of all source } \\ 
     &\text{mapped to $x$ to columns } \bar{c} \text{ where applicable}
\end{align*}

\item \textbf{external functions:}

\begin{align*}
     &\lambda \sigma^\sharp.I.\llbracket y = f(\bar{x}) \rrbracket  = \sigma^\sharp[ y \mapsto \sigma^\sharp(y) \sqcup \bigsqcup_{x \in \bar{x}} \sigma^\sharp(x)], I
\end{align*}

\item \textbf{sink}:
\begin{align*}
  &\lambda \sigma^\sharp.I.\llbracket m.f(\bar{x}) \rrbracket  =
    \sigma^\sharp, \forall x \in \bar{x}, I \cup \sigma^\sharp(x)\\
   & \text{ where } f \in \textit{Sink}
\end{align*}

\end{enumerate}

Case (1) handles read statements. Here we simply map the left-hand-side variable to the source being read with all columns. Case (2) handles projection. Here we map the left-hand-side variable to the source with appropriate constraints on columns. Case (3) handles external functions, here we simply join all sources from the arguments ($\bar{x}$) of the function to the left-hand-side variable. Case (4) handles a call to a model. Here we add the sources
of the arguments to $I$.  
\begin{example} [Static Analysis Example]
For \emph{Train1.py} in Figure~\ref{code:train1}, we proceed to process rule (1) for both read statements, finishing with the abstract state: 
$\{data1 \mapsto file1^{\dots, loc, age, \dots} , file2 \mapsto output^{\dots, target, \dots} \}$. 

The projection invokes rule (2) constraint so that we have variables $X$ and $y$ map to: 
$\{X \mapsto file1^{loc, age}, y \mapsto output^{target} \}$ 

Finally, at line~8, we apply rule (3) and when we detect the fit function in line~10 we apply rule (4) and add the 
data sources of  $X\_train$ and $y\_train$ to $I$, i.e., $I = \{ file1^{loc, age}, output^{target} \}$.
\end{example}

\subsection{Source Mapping Analysis for Other Artifacts}
Aside from scripts, other activities we support include database queries, multi-file scripts and notebooks. For example, queries can be converted to imperative  Intermediate Representations (IRs) (cf. Chapter 3 in~\cite{suboticthesis}) and subsequent CFGs analyzed like scripts in Section~\ref{sec:scripts}. In practice, we simply match the columns and tables on the AST, in similar spirit to~\cite{facebooksql}.  Multi-file scripts (e.g., pipelines) and functions can be resolved via 
cloning~\cite{cloning}. Notebooks can be handled by applying techniques in~\cite{SuboticMS22} which lift script static analyses to handle notebook semantics.

\section{Deployment and Applications}
In this section we describe the implementation of our mapping framework and  its deployment. We also discuss several use cases. 

\begin{figure}
\centering
\includegraphics[width=0.4\textwidth]{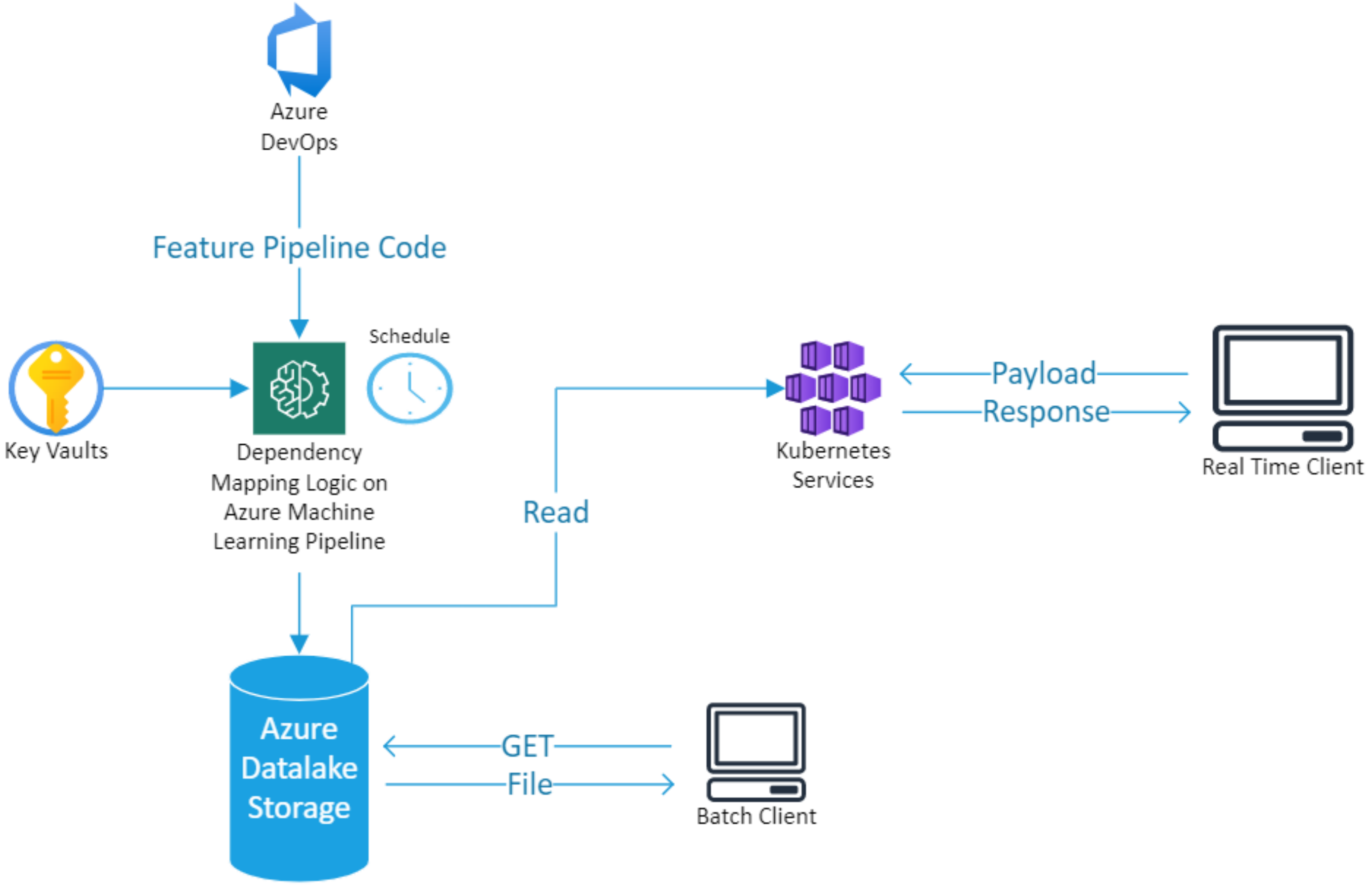}
\caption{High-level sketch of the software architecture implemented to expose dependency map as a RESP web API\label{fig:overview}.}
\end{figure}

The high-level overview of the architecture of the mapping framework API is presented in Figure~\ref{fig:overview}. We have deployed the dependency map as a REST web API service.  The request payload of the API consists of: (1) the path to an Azure DevOps Git repository hosting the target source code we wish to analyze and extract data 
dependencies from (2) An optional field to filter the response. The response consists of the dependencies of the models in the Git repository, filtered as per the optional field in the request. 
The overall logic is orchestrated within Azure Data Factory and can be scheduled to run either at a pre-defined schedule or following each pull request merges.
The code itself is containerized and the image contains the .NET runtime.  First, the target source code is 
pulled from Azure DevOps Git repository and processed by our mapping 
framework.  Finally, 
a copy of the data source dependency mappings is stored as a file in Azure Data Lake Storage (ADLS).  The service controllers are hosted separately on a Kubernetes cluster provided through Azure Machine Learning Inference Endpoints and the service pods cache the output file and serve the 
real-time requests.  
Below we describe three use cases that use our service. 

\paragraph{\textbf{Use Case I:} Interactive MLOps dashboards} Our first use case 
is an interactive PowerBI dashboard which hooks into the API for real-time updates.  The main purpose of the dashboard is to assist our MLOps engineers in maintaining good overall health of our production ML models.  For example, when MLOps engineers receive communication about data source changes, they rely on the dashboard to identify which activities may be impacted.  This way, we ensure that comprehensive and responsive actions are taken to react to data source changes.  In particular, the dashboard is useful in identifying non-trivial dependencies such as model-to-model dependencies.  Finally, the dashboard helps MLOps engineers decide if a model needs re-training in case some of its most important features are affected by a data source change.


\paragraph{\textbf{Use Case II:} Knowledge graphs} Traditional entity relationship diagrams quickly become outdated 
with ever changing data sources, model owners and downstream business stakeholders.  Recently, knowledge graphs have gained momentum as a way to collect information about complex datasets, their relationships to one another, connect business terms to data elements and more.  Additionally, knowledge graphs are a convenient tool to surface insights that improve data discovery and governance.  We have implemented our own version of a knowledge graph that leverages the dependency map API to cross-link between activities and their extended information.
Integrating this network of information with our Incident Management System (IcM) removes some of the manual load from MLOps engineers by sending automatic alerts to the relevant parties affected by a data source change.  Additionally, the dependency map API populates meta-information knowledge about the data sources to improve discoverability (such as column data types in tables, business stakeholders etc.).

\paragraph{\textbf{Use Case III:} Feature stores} Feature stores are a data management layer 
whose purpose is to act as a single source of truth of ML datasets.
Persisting features in a principled way is expected to accelerate model training and scoring so that compute-heavy features need not be re-evaluated every time they are requested.  Our dependency map API plugs into a feature store to identify feature re-usability and decrease the storage requirements even more.  For example, it is very common for multiple graphs to share a number of data sources.  This happens 
when different models re-use the same set of features.  The dependency map API  automatically populates the feature store meta-information so that data source duplication is explicitly taken into account in the design of the feature store persistence model.


\section{Experimental Evaluation}
\label{sec:evaluation}
In this section, we evaluate our implementation of our mapping framework. 

\paragraph{Experimental Setup}
We perform our evaluation based on $31$ real-world ML models\footnote{We provide IDs as model names are confidential.} that are leveraged by downstream stakeholder teams which rely on the output of these models to make more informed business decisions. 20\% of models require an inter-graph analysis.  Our experimental evaluation aims to validate the low latency of our system with respect to providing up to date alarms to model owners. All experiments were performed on an Azure Machine Learning workspace with Standard\_D12\_v2 compute configuration (26 GB RAM and 56 GB disk) running the Ubuntu 22.04.1 LTS operating system. 


\begin{table}[]
\caption{Dependency Map Evaluation}
\label{tab:perf}
\footnotesize
\begin{tabular}{|l|l|l|l||l|l|}
\hline
\textbf{Model ID}            & \textbf{Avg. Act Size} & \textbf{No. Act}        & \textbf{No. Dep}  & $\mathbf{|\zeta|}$    & \textbf{T (ms)}  \\\hline \hline
M1              & 121524           & 63                                     & 127             & 1591               & 742      \\
M2               & 199              & 1                                      & 37              & 1209               & 336      \\
M3                            & 173856           & 38                                     & 37              & 1145               & 986      \\
M4                     & 45482            & 18                                     & 47              & 978                & 302      \\
M5                    & 30008            & 12                                     & 31              & 446                & 126      \\
M6            & 19764            & 13                                     & 60              & 434                & 233      \\
M7                  & 22470            & 8                                      & 20              & 282                & 82       \\
M8            & 4279             & 6                                      & 59              & 206                & 72       \\
M9         & 4481             & 7                                      & 55              & 199                & 80       \\
M10                            & 7140             & 5                                      & 26              & 170                & 122       \\
M11                            & 13028            & 6                                      & 35              & 135                & 164      \\
M12                        & 39281            & 11                                     & 44              & 131                & 162      \\
M13                          & 1660             & 18                                     & 34              & 100                & 15       \\
M14                           & 22392            & 18                                     & 111             & 58                 & 599      \\
M15                           & 18139            & 8                                      & 55              & 39                 & 163      \\
M16                   & 1855             & 2                                      & 22              & 32                 & 87       \\
M17           & 1138             & 2                                      & 47              & 29                 & 24       \\
M18                   & 84               & 1                                      & 39              & 27                 & 22       \\
M19                & 2001             & 1                                      & 25              & 24                 & 53       \\
M20               & 4959             & 4                                      & 42              & 22                 & 81       \\
M21           & 2611             & 2                                      & 52              & 21                 & 76       \\
M22              & 22754            & 3                                      & 63              & 21                 & 136      \\
M23        & 682              & 1                                      & 23              & 20                 & 49       \\
M24            & 18140            & 8                                      & 40              & 20                 & 134      \\
M25               & 4847             & 3                                      & 26              & 19                 & 83       \\
M26                  & 2394             & 3                                      & 31              & 18                 & 59       \\
M27               & 4462             & 6                                      & 43              & 4                  & 719      \\
M28                   & 1497             & 2                                      & 42              & 4                  & 177      \\
M29                 & 142              & 1                                      & 35              & 3                  & 45       \\
M30                      & 168              & 2                                      & 65              & 2                  & 8        \\
M31  & 1494             & 4                                      & 25              & 2                  & 70       \\\hline \hline
\textbf{Geo. Mean} & 4630.2 & 4.8 &	40.8 & 50.3 &	107.8 \\\hline
\end{tabular}          
\end{table}

\paragraph{Experimental Results}
We present our experiments in Table~\ref{tab:perf}. Column \textbf{Avg. Act Size} measures the average size of each activity by number of tokens in the file. \textbf{No.Act} is the number of activities in the activity graphs, 
\textbf{No.Dep} is the number of dependencies (i.e., edges) in the activity graphs, 
\textbf{$|\zeta|$} is the number of (transitive and initial) data sources found to map to the model, 
\textbf{T} is the execution time (ms) to compute $\zeta$. We summarize each column by providing the geometric mean. 

Our experimental results show that all of our dependency mappings can be built in in under a second with a max. 986 ms and with a geo. mean of 107.8 ms. This conforms to execution times for similar static analyses on Python data science scripts~\cite{vamsa,SuboticMS22,SOAP}. We also observe that the size of the activities, the number of activities and the number of dependencies all have an influence on the execution time, while the size of $\zeta$ doesn't appear to have an influence. This conforms with our expectations as we statically analyse activity graphs and thus are unaffected by number of initial data sources. Overall, our system
is able to compute mappings after each pull request is merged so that dependencies are continuously kept up-to-date with respect to the SLAs we typically encounter. 

\section{Related Work}
\label{sec:rel}
The management~\cite{amazon} and development~\cite{SEforML} of ML 
systems is a rapidly emerging research area. Amongst the body of works in this area,
our technique most resembles methods that extract information from models using 
provenance/lineage. For instance, the technique in~\cite{mlinspect} provides run-time based 
lineage of ML pipelines. The technique in~\cite{deptrack} performs run-time tracking and uses block-chain systems 
to mitigate risk. Mistique~\cite{mistique} collects intermediate execution information and allows 
engineers to query for diagnosis of faults. ModelDB~\cite{ModelDB} stores 
trained models to enable querying of metadata and artifacts by exposing a 
logging API for a specific set of libraries. In contrast to these systems,
our system focuses on quickly mapping dependencies between data sources and 
models by leveraging static analysis. Since we do not require
a fine-grain data (actual values) run-time diagnostics provide little benefits and results in unnecessary burdens such as execution logging. Our technique on the other hand, requires no code modifications, making integration of activities seamless.  

We are unaware of other work that performs static analysis on activity graphs or similar structures. In terms of techniques that leverage static analysis more broadly for ML~\cite{SuboticMS22,SOAP}, 
our technique has similarities with Vamsa~\cite{vamsa}. Vamsa builds a static provenance 
directed acyclic graph (DAG) from a single Python script using a forwards/backwards propagation on acyclic control-flow programs. Compared to Vamsa, we analyze activity graphs which may contain various connected code artifacts including scripts, pipelines and queries. Our static analysis also has foundational similarities with the dependency analysis in~\cite{arxiv.2211.16073}, which could be used as an intermediate semantics to prove soundness. 
\section{Conclusion}
\label{sec:conclusion}
We have presented a 
dependency mapping framework for large scale 
ML models that are defined by inter-connected activity graphs. Our technique statically analyzes these activity graphs to compute their initial data source dependencies.
We have deployed our framework within MCDS as a service that is leveraged in a number of use cases. In the future we plan on expanding our coverage of ADF activities and investigate the use of static analysis  for other MLOps use cases. 
  





\begin{acks}
We thank our colleagues in Microsoft Cloud Data Sciences and Azure Data Labs for their feedback and support. In particular, we thank Kiran R, Kirk Li, Swarnim Narayan, Rituparna Praharaj and Guillaume Bou\'e for their valuable comments and help.
\end{acks}

\bibliographystyle{ACM-Reference-Format}
\bibliography{main}

\end{document}